\documentclass[authoryear]{elsarticle}

\usepackage{natbib}

\usepackage{graphics}

\usepackage{amssymb}

\begin{document}

\begin{frontmatter}



\title{Constraints on a hadronic model for unidentified off-plane galactic gamma-ray sources}


\author{Wilfried F. Domainko}
\address{Max-Planck-Institut f\"ur Kernphysik\\
Saupfercheckweg 1\\
69115 Heidelberg, Germany}
\ead{wilfried.domainko@mpi-hd.mpg.de}

\begin{abstract}

Recently the H.E.S.S. collaboration announced the detection of an unidentified gamma-ray source with an off-set from the galactic plane of 3.5 degrees: HESS J1507-622. If the distance of the object is larger than about one kpc it would be physically located outside the galactic disk. The density profile of the ISM perpendicular to the galactic plane, which acts as target material for hadronic gamma-ray production, drops quite fast with increasing distance. This fact places distance dependent constraints on the energetics and properties of off-plane gamma-ray sources like HESS J1507-622 if a hadronic origin of the gamma-ray emission is assumed. For the case of this source it is found that there seems to be no simple way to link this object to the remnant of a stellar explosions.

\end{abstract}


\end{frontmatter}

\parindent=0.5 cm

\section{Introduction}

The H.E.S.S. collaboration has performed a survey along the galactic plane in very-high energy (VHE, $>$100 GeV) gamma rays \citep[e.g.][]{aharonian06}. With this program several VHE sources have been discovered and it has further been found that these objects are concentrated at the galactic plane \citep[e.g.][]{chaves09}. Some of these sources lack counterparts in other wavelength \citep{aharonian08}. Recently, a source with no obvious counterpart in other wavelength has been found with a significant off-set of 3.5 degree from the galactic plane: HESS J1507-622 \citep{tibolla09a, tibolla09b}. This source is detected with an integral flux ($>$ 1 TeV) of $(1.5 \pm 0.4_{stat} \pm 0.3_{sys} ) \times 10^{-12}$ cm$^{-2}$ s$^{-1}$ and a spectrum given by $dN/dE = k (E/1 \mathrm{TeV})^{-\Gamma}$ with $\Gamma = 2.24 \pm 0.16_{stat} \pm 0.20_{sys}$ and a flux normalization of $k = (1.8 \pm 0.4) \times 10^{-12}$ TeV$^{-1}$ cm$^{-2}$ s$^{-1}$ \citep{tibolla09b}. The angular distance of HESS~J1507-622 to the plane results in some interesting constraints on the nature of the source. Either it is rather nearby or
if the distance of the object is larger than about one kpc it would be also physically located outside the galactic disk. Its rather compact angular size in comparison to the apparent extension of other VHE sources would suggest a multiple-kpc distance \citep{hinton09}.

VHE gamma-rays can be generated mainly by two different channels: through inelastic collisions from hadronic cosmic rays with thermal target protons and nuclei and subsequent $\pi^0$ decay (hadronic channel) or by inverse Compton up-scattering from low energy photons by high energy electrons (leptonic channel). Additional mechanisms which can produce VHE gamma-rays include interactions of cosmic ray protons with low energy photons and subsequent $\pi^0$ production and decay and photo-disintegration of heavy nuclei \citep[see][]{aharonian04}. For HESS J1507-622 the gamma-ray emission scenario is not clear at the moment. For the purpose of this paper a hadronic origin is assumed. This scenario seems to be more plausible due to the absence of a clear low-energy counterpart. A hadronic model involving interactions between high-energy protons with the ambient inter-stellar medium (ISM) is adopted to investigate the influence of a location at low density environments on the properties of such sources. The gamma-ray luminosity in this picture depends on the density of target material which is monotonically decreasing with increasing distance from the galactic plane \citep{dickey90}. 

In many cases the sources of hadronic VHE gamma rays will be linked to the remnants of catastrophic events at the end of stellar evolution.
Objects with significant physical distances from the galactic plane are expected to belong to an old stellar population. Supernovae (SN) type Ia are typical representatives for stellar explosions connected to such a stellar population. A VHE gamma-ray emission from the remnant of SN 1006, which was of type Ia, has already been detected \citep{naumann08,acero10}. This remnant is located at a distance from the galactic plane of 500 pc in an environment with a very low density of target material of 0.085~cm$^{-3}$ \citep{acero10}. Further sources of gamma-rays with potential off-plane locations might be the remnants of compact binary mergers \citep{domainko05,domainko08} and the remnants of low luminosity GRBs \citep{foley08}. The recent discovery \citep{badenes09} of a nearby close white dwarf - neutron star binary \citep[but see][for a different interpretation]{marsh10} and the high rate of mergers of these objects deduced from this discovery \citep{thompson09} might represent an additional channel for producing gamma-ray sources connected to an old stellar population.

\section{Energetics}
\label{energetics}

\subsection{Large scale ISM distribution as ambient medium}

In a hadronic gamma-ray production scenario the gamma-ray luminosity scales linear with the energy in cosmic rays and the density of the target material. Objects without strong stellar winds like white dwarf binaries should not alter their environment and therefore the density of target material is given by the ambient ISM. For off-plane sources the density of target material is expected to drop monotonically with distance since larger distances would place the object at larger physical off-sets from the Galactic disk. In the direction of HESS J1507-622 there is some evidence of molecular gas at a distance of 400 pc but with no clear connection to the gamma-ray source \citep{tibolla09b}. At larger distances than this there is no other indication for further clouds seen. Therefore, for the purpose of this section, to estimate the density of target material the galactic density profile perpendicular to the galactic disk of \citep{dickey90} is used. Since HESS J1507-622 is located at a galactic latitude  of about 317$^\circ$ the closest galactocentric distance of the line of sight is about 5.8 kpc and for distances smaller than 10 kpc its galactocentric distance will always be in the range of 4 -- 8 kpc. Hence the best fitting model of the ISM density in this galactocentric distance range of \citep{dickey90} is assumed to obtain the distance dependent density of target material along the line of sight for HESS J1507-622. In this framework this is a conservative approach since this model gives the highest density of all estimates up to a distance of about 5 kpc in the direction of the VHE source \citep[see Fig. 10 of][]{dickey90}. The best fitting model is given by

\begin{equation}
n(z) = 0.395\, e^{-z^2 /127 \mathrm{pc}} + 0.107\, e^{-z^2 /318 \mathrm{pc}} + 0.064\, e^{-z / 403 \mathrm{pc}}\, \mathrm{cm^{-3}}
\end{equation}

To calculate the number density $n(z)$ at the position of HESS~J1507-622 as a function of distance $d$ a physical off-set from the disk $z$ of $z = d \times sin(3.5^\circ)$ is used.
The density of target material as a function of distance is plotted in Fig.1. 

The energy in cosmic rays $E_\mathrm{CR}$ needed to explain the flux of the VHE source is calculated according to:

\begin{equation}
E_\mathrm{CR}(d) \approx F_{\gamma}(>1\, \mathrm{TeV}) \times 4 \pi d^2 \times \tau_\mathrm{pp} \times \eta^{-1} \times (1+S)
\end{equation}

Here $F_{\gamma}(>1\, \mathrm{TeV})$ is the integral gamma-ray flux above an energy of 1 TeV, $\tau_\mathrm{pp}$ is the cooling time of hadronic cosmic rays (given by $\tau_\mathrm{pp} = (n(d) \sigma_{pp} f c)^{-1}$ with $n(d)$ is the distance dependent number density, $\sigma_{pp}$ is the proton - proton cross section of about 40 mb, $f \approx 0.5$ is the in-elasticity parameter and $c$ is the speed of light \citep{aharonian04}), $\eta \approx 0.3$ is the fraction of the energy of the hadronic interaction channeled into gamma rays and $S$ accounts for cosmic rays which produce gamma rays below 1 TeV ($S \approx 5$, for a cosmic ray spectral index $\Gamma$ = 2 down to a low energy cut-off of the cosmic ray energy of 1 GeV).
The energy in cosmic rays as a function of distance can again be seen in Fig. 1.
A spectral energy distribution (SED) for a hadronic model is shown in Fig. 2.

A typical energy of 10$^{50}$ ergs produced by a SN would place HESS~J1507-622 below a distance of about 2 kpc. If the source would contain 10$^{51}$ ergs in hadronic cosmic rays it would be located at a distance of 4 kpc. Since relativistic shock waves are expected to convert a considerable fraction of their initial energy into cosmic rays \citep{atoyan06} a hypothetical GRB remnant could be located roughly at such a distance. For distances larger than 5 kpc the energy requirements are getting unrealistically large. This results from a superposition of increased distance and decreased ambient density. For an increase in distance from 2 kpc to 6 kpc the energy in cosmic rays will increase by a factor of 40. The contribution to this factor due to the increased distance is a factor 9 and the contribution of the decreased density is about 4.4.

To summarize, the main results from this section are that a typical energy in cosmic rays produced by a supernova of 10$^{50}$ ergs would place HESS~J1507-622 at a distance of less than 2 kpc while distances to this source of up to 5 kpc would still be possible for exceptionally energetic events.

\subsection{Stellar wind blown bubbles as target material}

High mass stars typically exhibit strong stellar winds and can therefore create region with enhanced density around them. In general massive stars are usually found close to the galactic plane but runaway stars dynamically ejected from young clusters can travel to off-plane locations \citep[e.g.][]{dewit05}. In this section it is explored how a progenitor star which featured a strong stellar wind in the past can alter the properties of the embedding medium into which it explodes as a SN.

The number density of the wind material as a function of distance to the star $n(r)$ can be expressed by:

\begin{equation}
n(r) \approx \dot M / (m_\mathrm{p} V_\mathrm{shell}(r))
\end{equation}

with $\dot M$ is the mass loss rate of the star, $m_\mathrm{p}$ the mean particle mass in the wind and $V_\mathrm{shell}(r) \approx 4 \pi r^2 s$ is the volume of a shell at radius $r$ from the star and a thickness $s$ which the wind traverses with its velocity $v$ in unit time. The angular radius of HESS J1507-622 is 0.15$^{\circ} \pm 0.02^{\circ}$ \citep{tibolla09b} and consequently $r(d) = sin(0.15^{\circ} \times d)$ with $d$ being the distance to the source. The number density of a massive stellar wind ($\dot M = 10^{-3}\, \mathrm{M}_\odot$/yr, $v = 500$~kms$^{-1}$) at the extension of the VHE source is plotted in Fig.3 as a function of distance. It can be seen that even for such a conservative case the number density of the embedding medium of HESS J1507-622 would drop to the level of the large scale ISM distribution at a distances of about 2 kpc. The density of a wind-created ambient medium is also compared to the density of target material which would be required if the energy in cosmic rays is fixed for all distances to 10$^{50}$ ergs (see Fig. 3). It is found that a stellar wind can only provide a high enough number density of target material for distances smaller than about 2 kpc for such a case.

\section{Age of the object}

The age of the hypothetical remnant can be estimated according to its extension. There are two different ways how the age of a remnant is linked to its extension. In the first model it is assumed that the remnant follows the typical evolution of a SNR which starts with free expansion and then, after the ejecta has pushed away a comparable mass of the ISM, enters the pressure driven Sedov phase. In this scenario the size of the gamma-ray source corresponds to the expanding shell of the remnant. In the second scenario it is adopted that cosmic rays were injected in a point-like explosion and the extension results from diffusion of the cosmic rays away from the location of the injection \citep[e.g.][]{atoyan06}.

\subsection{SNR evolution}

For a SNR the transition time $t_\mathrm{f}$ between the free expansion phase and the Sedov phase is given by:

\begin{equation}
t_\mathrm{f} \approx \frac{1}{v_\mathrm{ej}}\, \sqrt[3]{\frac{3\, M_\mathrm{ej}}{4 \, \pi \, n \, m_\mathrm{p}}} 
\end{equation}

with $v_\mathrm{ej}$ the velocity and $M_\mathrm{ej}$ the mass of the ejecta, $n$ the number density and $m_\mathrm{p}$ the mean particle mass of the ambient medium. For the case of a SN Ia ($M_\mathrm{ej} = 1.4\, \mathrm{M_\odot}$, $v_\mathrm{ej} = 12 000\,$km~s$^{-1}$, E$_\mathrm{kin} = 10^{51}\,$ergs) at a distance of 2 kpc and thus a number density given by the large scale ISM distribution of 0.3 cm$^{-3}$, the transition time $t_\mathrm{f}$ would be about 290 years. For a massive progenitor star with $M_\mathrm{ej} = 10\, \mathrm{M_\odot}$, $v_\mathrm{ej} = 3000\,$km~s$^{-1}$ (E$_\mathrm{kin} = 10^{51}\,$ergs) located in the same ambient medium at the same distance the transition time would be about 2250~years.

To estimate the age according to the Sedov expansion it is assumed that 10\% of the initial kinetic energy of the shock wave $E_\mathrm{tot}$ is channeled into cosmic rays equivalent to $E_\mathrm{tot}(d) = 10 \times E_\mathrm{CR}(d)$ where $E_\mathrm{CR}(d)$ is the distance dependent energy in cosmic rays from Sec. \ref{energetics}. It is further assumed that the remnant evolves in an ambient medium with the distance dependent density $n(d)$ from Sec. \ref{energetics}. The age according to the Sedov solution $t_\mathrm{sedov}$ is:

\begin{equation}
t_\mathrm{sedov} \approx \sqrt{\frac{n(d) \, m_p \, r(d)^5}{2 \, E_{tot}(d)}}
\end{equation}

Here $r(d)$ is the distance dependent radius of the remnant. For a distance of 2 kpc and the corresponding $E_{CR}$ of about $2 \times 10^{50}\,$ergs from section \ref{energetics} the Sedov age would be about 340 years. This is larger than the transition time for the case of a SN Ia but smaller than this time scale for a massive progenitor. Thus a remnant of a SN Ia would be already in the Sedov phase whereas the remnant of a massive progenitor would still be freely expanding. A remnant with the observed extension of HESS~J1507-622 at a distance of 2 kpc resulting from the massive progenitor scenario with $v_\mathrm{ej} = 3000\,$km~s$^{-1}$ would have an age of about 1650 years. Since SNRs are considered to be efficient accelerators of cosmic rays at the beginning of the Sedov phase the SN Ia scenario with the pressure driven age is considered here.
The age of a hypothetical pressure driven remnant as a function of distance with the extension of HESS J1507-622 can be seen in Fig.~4.

A remnant which produced the typical SN energy in cosmic rays of $10^{50}$~ergs at a distance of about 2 kpc would have a pressure driven age of less than 1000 years. Thus it seems that a SN Ia remnant interpretation of HESS~J1507-622 would only be possible for a remnant with similar age as SN 1006. Since SN 1006 is a prominent source in a wide multiwavelength range it would be surprising for such an interpretation that HESS~J1507-622 lacks comparable signatures. In general, the Sedov age for HESS~J1507-622 for any given distance never exceeds about 1000 years. This can be understood by the fact that for larger distances larger energetics are involved (see Fig. 1) and therefore the hypothetical remnant expands accordingly faster.

\subsection{Age of a hypothetical remnant according to the diffusion scenario}

For the case of a GRB remnant it is assumed that cosmic rays were accelerated in the relativistic shocks created by the central engine \citep{atoyan06}. After $\lesssim$ 1 year these shocks will become non-relativistic and the cosmic rays which were accelerated in the relativistic shock waves will diffuse away from the injection site. Hence for time-scales $\gg$~1~year the evolution of such a remnant can be described by cosmic ray diffusion following point-like injection.
The age of a remnant $t_\mathrm{diff}$ with extension $r(d)$ determined by diffusion of cosmic rays can be expressed by:

\begin{equation}
t_\mathrm{diff} \approx r(d)^2/2\, D(E)
\end{equation}

with $D(E) = 0.3 \times 10^{27} (E/10 \, \mathrm{GeV})^{0.5}$ cm$^2$ s$^{-1}$ is the energy dependent diffusion coefficient \citep{atoyan06}. It has to be noted that this diffusion coefficient was obtained for locations in the galactic plane and that the diffusion coefficient may differ at off-plane locations from this value. With the galactic diffusion coefficient and adopting the source extension of 0.15$^{\circ}$ an age of the remnant of $t_\mathrm{diff} \lesssim 200\, d^2_\mathrm{kpc}$ years is found \citep{atoyan06}. The distance dependent age of such a remnant can again be seen in Fig. 4.

For a remnant which size is defined by the diffusion of cosmic rays like it was proposed for GRB remnants, the age will be smaller than $10^4$ years for distances which would involve reasonable energetics ($\lesssim 5 \times 10^{51}$ ergs) and thus are smaller than 5 kpc. 
This diffusion based age estimate can be shorter than the galactic rate of rare events. For example
the galactic neutron star - neutron star merger rate was estimated to one event per (0.5 - 7)$\times 10^4$ years \citep{kalogera04}.
Therefore such a scenario appears to be possible but rather unlikely.

To summarize, in a scenario where the extension of the remnant is determined by cosmic ray diffusion the rather compact size of HESS~J1507-622 would favor a reasonable frequent astrophysical event (galactic rate $\gtrsim$ 10$^{-4}$ per year) as its origin.

\section{Discussion}

With the example of HESS~J1507-622 it has been shown how constraints could be placed on a hadronic model for off-plane gamma-ray sources if the VHE emission is generated by the remnants of stellar explosions. For HESS~J1507-622 it has been found that only a young SNR ($\lesssim$ 1000 years) could explain the extension of the source if it is linked to the size of the entire remnant. For such a scenario the absence of prominent multiwavelength counterparts seem to be puzzling. 
Especially X-ray emission produced by synchrotron radiation of primary electrons accelerated in the same blast wave as the hadronic cosmic rays would be expected. From limits on the X-ray flux constraints on the electron to proton ratio can be obtained if a magnetic field strength is assumed. For a conservative magnetic field strength of 3 $\mu$G and using the flux of the closest source from the ROSAT all-sky survey \citep[1RXS J150841.2-621006,][]{voges00} as a comparison it is found that the electron to proton ratio should not significantly exceed 10$^{-5}$ (see Fig. 2). However, it has to be noted that the caveat for such a comparison is the different extension of the VHE and the X-ray source. For a higher upper limit on any diffuse X-ray emission the constraints on the electron to proton ratio is more relaxed. Higher values for the magnetic field would constrain the ratio of electrons to protons to even lower limits.

To summarize, a typical value for the energy in cosmic rays for a supernova of 10$^{50}$ ergs would place HESS~J1507-622 at a distance of less than 2 kpc. Such a distance constraint would require a pressure driven age of the remnant of less than 1000 years. For such a young remnant a small electron to proton ratio of 10$^{-5}$ would be surprising. 
Also for the case of a hypothetical GRB remnant the distance of the source would be limited by constraints on the energetics to $\lesssim$~5~kpc. In this scenario
the size of the remnant would result from the diffusion of cosmic rays away from the explosion site and the remnant would be relatively young ($\lesssim 10^4$ years). This might be difficult to explain if the remnant was created by a rare event like a compact binary mergers.

\section*{Acknowledgements}

The author acknowledges support from his host institution. The use of M. de Naurois modelisation package is acknowledged. The author thanks the referees for several interesting suggestions which improved the paper.

\newpage

\begin{figure}
\label{ecr}
\begin{center}
\includegraphics*[width=9cm]{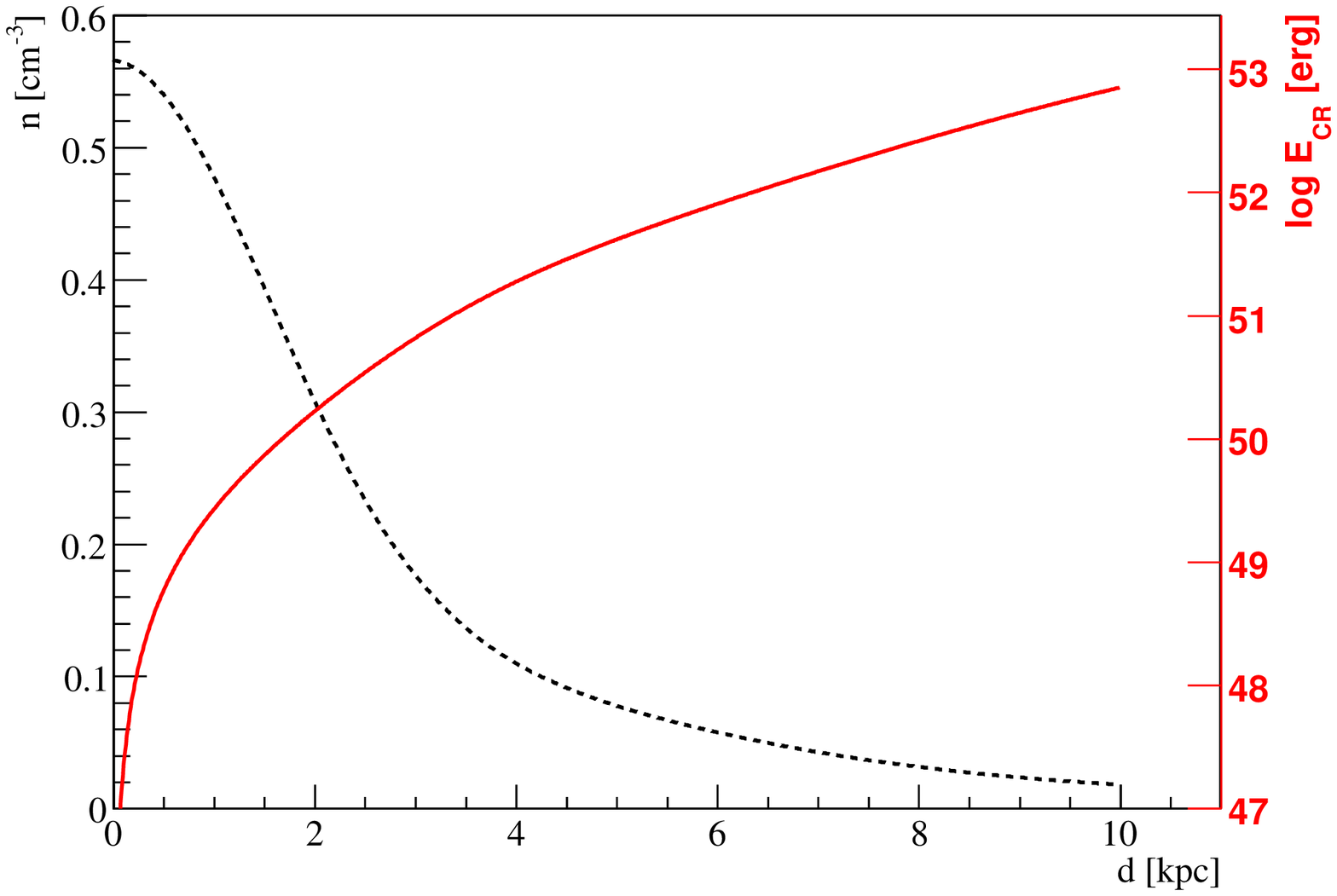}
\end{center}
\caption{The density of target material for hadronic gamma-ray production at the location of HESS~J1507-622 as a function of distance is plotted with a black dashed line. With the red solid line the energy in hadronic cosmic rays needed to explain the luminosity in VHE gamma radiation again as a function of distance is illustrated.}
\end{figure}

\begin{figure}
\label{sed}
\begin{center}
\includegraphics*[width=10cm]{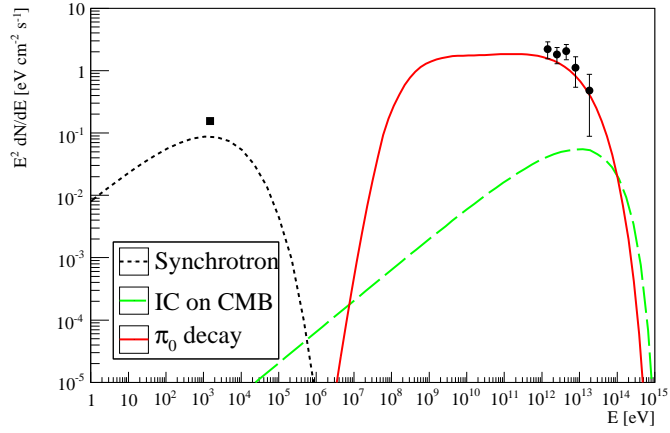}
\end{center}
\caption{A calculated SED for a hadronic model with $d = 2$ kpc, $n = 0.3$ cm$^{-3}$, E$_\mathrm{CR} = 2 \times 10^{50}$ erg and spectral index $\Gamma = 2$ between 1 GeV and 100 TeV is shown. Additionally, leptonic signatures from primary electrons for a single zone model with the same spectrum, assuming a magnetic field of 3 $\mu$G and an electron to proton ratio of 10$^{-5}$ is plotted. The measured flux in the VHE gamma-ray range obtained with H.E.S.S. and, for comparison, the X-ray flux of the faint ROSAT all-sky survey source 1RXS J150841.2-621006 are also displayed.}
\end{figure}

\begin{figure}
\label{wind}
\begin{center}
\includegraphics*[width=9cm]{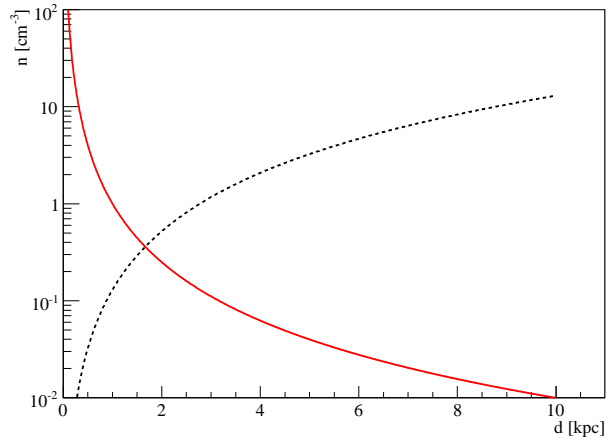}
\end{center}
\caption{The density of the ambient medium at the extension of HESS~J1507-622 provided by a massive stellar wind of a progenitor star ($\dot M = 10^{-3}\, \mathrm{M}_\odot$/yr, $v$ = 500 kms$^{-1}$) as a function of distance is plotted with a red solid line. For comparison the required density of target material for the case where the energy in cosmic rays is fixed to 10$^{50}$ ergs for all distances is shown with the dashed black line. A stellar wind can only provide a high enough number density of target material for distances smaller than about 2 kpc for such a case.}
\end{figure}

\begin{figure}
\label{age}
\begin{center}
\includegraphics*[width=9cm]{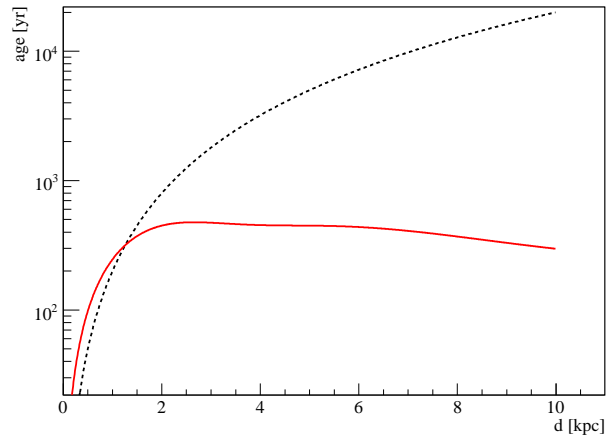}
\end{center}
\caption{The age of a pressure driven remnant in the Sedov phase with the angular extension of HESS~J1507-622 as a function of distance is plotted with the red solid line. With the black dashed line the age of a remnant that's extension is defined by the diffusion of cosmic rays after a point-like injection is plotted. For details see main text.}
\end{figure}

\end{document}